\documentclass{aipproc}

\layoutstyle{6s}

\newcommand{\Bsat}{B_{\rm sat}}
\newcommand{\e}{\varepsilon}
\newcommand{\gn}{\gamma_{b0}}
\newcommand{\kpeak}{k_{\rm peak}}
\newcommand{\me}{m_\rme}
\newcommand{\mi}{m_\rmp}
\newcommand{\ompb}{\omega_{\rm pb}}
\newcommand{\ompe}{\omega_{\rm pe}}
\newcommand{\ompp}{\omega_{\rm pi}}
\newcommand{\overgbar}{\left(\overline{1\over\gamma_b}\right)}
\newcommand{\pxn}{p_{x0}}
\newcommand{\px}{p_x}
\newcommand{\py}{p_y}
\newcommand{\pzn}{p_{z0}}
\newcommand{\pz}{p_z}
\newcommand{\rmd}{{\rm d}}
\newcommand{\rme}{{\rm e}}
\newcommand{\rmp}{{\rm p}}
\newcommand{\uxn}{u_{x0}}
\newcommand{\uzn}{u_{z0}}
\newcommand{\va}{v_{\rm A}}
\newcommand{\vecp}{\vec{p}}
\newcommand{\vxn}{v_{x0}}
\newcommand{\vzn}{v_{z0}}

\title{Magnetic field generation in relativistic shocks}

\author{Jorrit Wiersma}{address={Sterrenkundig Instituut, Universiteit
Utrecht, Netherlands}}
\author{A. Achterberg}{address={Sterrenkundig Instituut, Universiteit
Utrecht, Netherlands}}

\begin{abstract}
We present an analytical estimate for the magnetic field strength generated
by the Weibel instability in ultra-relativistic shocks in a hydrogen plasma.
We find that the Weibel instability is, by itself, not capable of
converting the kinetic energy of protons penetrating the shock front
into magnetic field energy.
Other (nonlinear) processes must
determine the magnetic field strength in the wake of the shock.

\smallskip\noindent
The following article has been submitted to 2003 GRB Conference
Proceedings. After it is published, it will be found at
\url{http://proceedings.aip.org/}.
\end{abstract}

\begin{document}

\maketitle

\section{Introduction}

The fireball model for Gamma-ray Bursts (GRBs) explains GRB afterglows as
synchrotron radiation coming from the external shocks that are formed when
a relativistically expanding fireball (or jet) interacts with surrounding
gas~\cite{reme92}.
The spectra and luminosity of the observed afterglows indicate a magnetic
field strength in the radiating material of about 10\% of the equipartition
field strength~\cite{grwa99}: $B^{2}/8 \pi \sim \e$ with $\e$ the total
post-shock energy density.
Such a magnetic field is much stronger than what is expected from simple
shock physics: for instance, passive compression of a dynamically
unimportant magnetic field in a relativistic shock yields a post-shock
field satisfying (for example, see~\cite{keco84}) $B^{2}/8 \pi \e \sim
(\va/c)^{2}$, with $\va$ the Alfv\'en speed {\em ahead} of the shock, which
usually satisfies $\va \ll c$.

It may be possible that much stronger magnetic fields are generated in the
shock transition itself through a Weibel-like instability.
This low-frequency electromagnetic beam-instability can develop at the
point where the shocked and unshocked plasma penetrate each other.
It converts the kinetic energy of the penetrating particle beams into
thermal motions and unordered magnetic fields~\cite{yal94}.
For external shocks propagating into a hydrogen plasma incoming protons
carry most of the kinetic energy in the shock frame.
The instability must convert a significant fraction of this energy into
magnetic energy in order to reach the required field strength~\cite{melo99}.
Here we investigate if this is indeed possible.

\section{Simple collisionless shock model}

Within the shock transition, where the fireball material encounters the
surrounding plasma, the mixing of the shocked (relativistically hot) and
the unshocked (cold) plasma produces a plasma with a very anisotropic
velocity distribution.
As in non-relativistic collisionless shocks, it is quite likely that a
significant fraction of the incoming ions is reflected at the shock
transition by a large-scale electrostatic or magnetic field.
The resulting situation is known to be unstable~\cite{yal94}: in the
so-called Weibel instability the penetrating and reflected particles will
bunch together, causing electric currents that induce a magnetic field in
the plasma, which in turn causes the particles to bunch even more.
If the generated magnetic fields become sufficiently strong they will trap
the beam particles, eventually saturating the instability.

Recent numerical simulations~\cite{fstm03}, \cite{fhhn03},~\cite{hasa03}
show that the electric currents generated by the Weibel instability
merge with each other in the wake of the unstable region.
This separate process has a longer time-scale, and forms structures
on a larger length-scale than the Weibel instability, which occurs mostly
on a scale of the order of the plasma skin depth $c/\ompe$, with
$\ompe$ the electron plasma frequency (see below).

In this paper we will concentrate on the proton-driven Weibel instability.

\section{The Weibel Instability and its saturation}

Because of the small electron mass, the electron-driven Weibel instability
evolves very rapidly~\cite{melo99}, with the magnetic field strength
growing as $e^{\sigma t}$ with $\sigma \approx \ompb$ where $\ompb =
    \sqrt{4 \pi e^{2} n_{0 \rm b}/\me}$ is the beam particle plasma
    frequency based on the proper beam density~$n_{0\rm b}$.
When the electron-instability has saturated, the electron velocity
distribution will be close to isotropic.
The electrons then form a relativistically hot background plasma in which
the much slower proton-driven Weibel instability develops because the
proton velocity distribution is still very anisotropic (see, for example,
figure~6 of~\cite{fhhn03}).
We will investigate the behavior of the protons in this situation.

The main features of the resulting instability can be reproduced by
investigating a water-bag proton velocity distribution.
We assume two counter-streaming proton beams moving along the
$x$-direction, with a small velocity spread in the $z$-direction to model
thermal motions.
The proton momentum~$\vecp$ is then distributed as:
\begin{equation}
    F(\vecp) = {n_\rmp\over 4 \pzn}[\delta(\px-\pxn) + \delta(\px + \pxn)]
	\delta(\py) [\Theta(\pz-\pzn) - \Theta(\pz+\pzn)],
\end{equation}
where $n_\rmp$ is the total proton density, $\pxn$ is the beam momentum,
$\pzn$ is the maximum momentum in the perpendicular direction, $\delta(x)$
is the Dirac delta function and $\Theta(x)$ is the unit step function.
We consider the evolution of a wave perturbation with wave vector $\vec{k}
= k \vec{e}_z$ and frequency~$\omega$.
The growth rate of the proton instability is calculated in the usual manner
by looking for wave solutions of the linearized equations of motion and
Maxwell's equations (for example~\cite{sftm02}).
The resulting dispersion relation links the complex wave frequency $\omega$
to the wave number $k$.
The Weibel instability obeys a dispersion equation of the form
(using the notation of~\cite{sftm02}):
\begin{equation}
    \omega^2 - k^2 c^2 + C_{xxz} = 0,
    \label{eq:deq}
\end{equation}
where $C_{xxz}$ contains contributions from both the electrons and the
protons.
Since the electrons are relativistically hot, their contribution is
$C_{xxz,\rme} = -\ompe^2$, with $\ompe$ the effective electron plasma
frequency which equals
\begin{equation}
    \ompe = \sqrt{4 \pi e^{2} n_{\rme}/\me h} \; ,
\end{equation}
where $h \equiv (\e +P)_{\rme}/n_{\rme} \me c^{2}$ is the electron enthalpy
divided by the rest mass energy density, which parameterizes the
relativistic mass correction in a relativistically hot plasma.

The proton contribution is~\cite{sftm02}:
\begin{equation}
    C_{xxz,\rmp} = \ompp^2 \left\{ -\overgbar
    +{1\over\gn}{\uxn^2\over 1+\uxn^2}
    - {1\over\gn}{k^2\vxn^2\over\omega^2-k^2\vzn^2}\right\},
\end{equation}
where $\ompp= \sqrt{4 \pi e^{2} n_\rmp/\mi}$
is the (non-relativistic) proton plasma frequency based on the lab-frame beam density,
$m_\rmp$ is the proton rest mass,
$u_i = p_i / (m_\rmp c)$, $\gn = (1 + \uxn^2 + \uzn^2)^{1/2}$, $v_i =
u_i c/\gn$ and
\begin{equation}
    \overgbar = \int\rmd\vecp\,{F(\vecp)\over\gamma}
    = {1\over 2\uzn} \: \ln\left( 1 + \vzn/c \over 1 - \vzn/c \right).
\end{equation}

\begin{figure}
\centerline{%
    \includegraphics[height=1.7in]{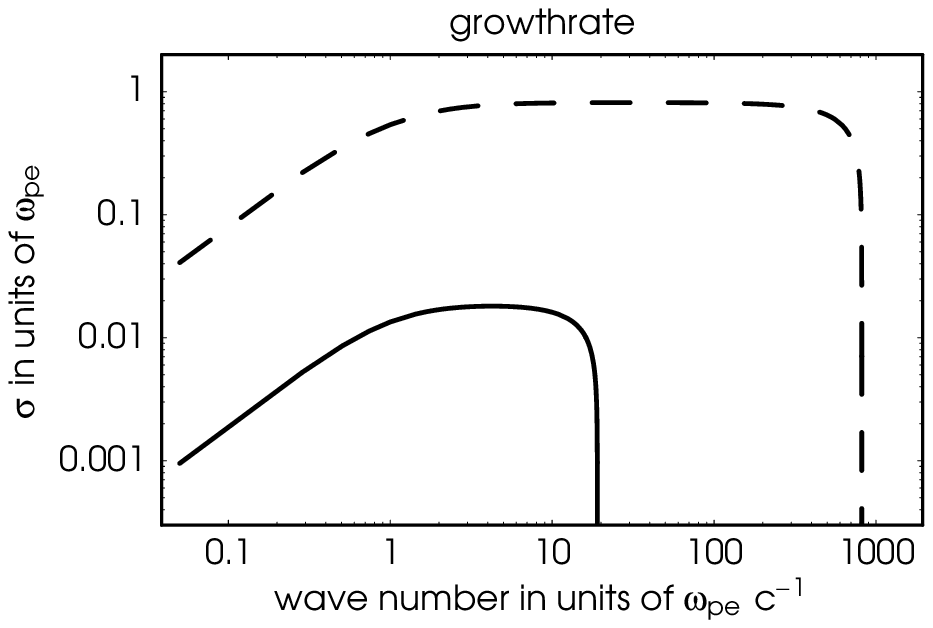}\enspace
    \includegraphics[height=1.7in]{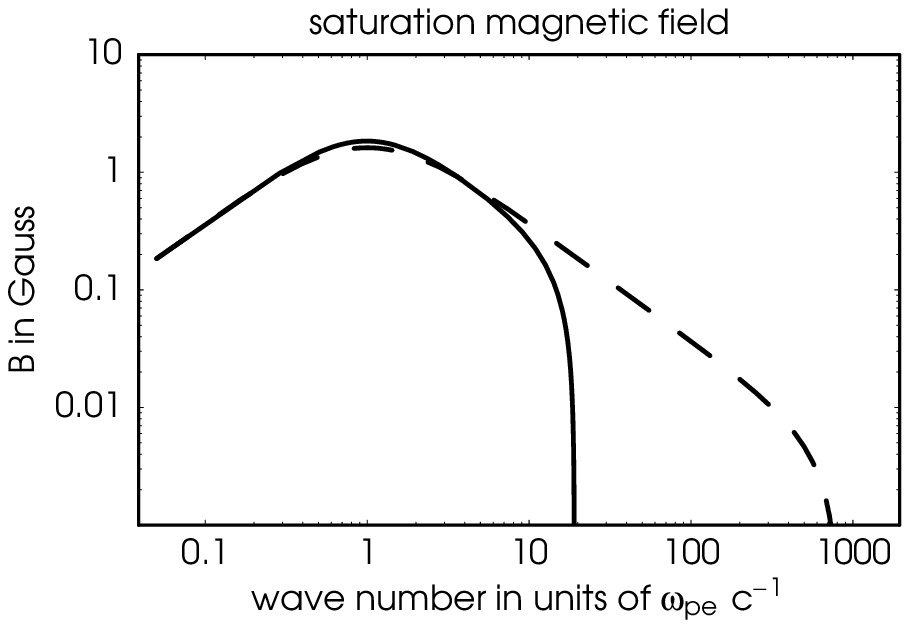}}
\caption{\label{fig:bsat}
Left: growth rate as a function of wave number for a shock with Lorentz
factor~$1000$ and thermal velocity spread~$\vzn = 0.001$.
The solid line is the result for an electron-proton plasma.
The dashed line is the result for pure pair plasmas.
Right: magnetic field strength as a function of wave number for a shock
with the same parameters and~$n_\rmp = 2\,{\rm cm}^{-3}$.}
\end{figure}

With these results the dispersion equation~(\ref{eq:deq}) becomes a
biquadratic equation for~$\omega$ that has one positive imaginary solution
giving the growth rate~$\sigma = {\rm Im}(\omega)$ of the unstable mode
(figure~\ref{fig:bsat}, left).

One expects the instability to saturate when the quiver motion of the beam
particles in the wave reaches an amplitude $\Delta z$ such that $k \Delta z \approx 1$.
An expression for the corresponding magnetic field strength is derived
in~\cite{yal94} and reads in our notation:
\begin{equation}
    	\Bsat = {\gn \,\mi \over \vxn \,e}{\sigma^2\over k}.
	\label{eq:bsat}
\end{equation}
Using the dispersion relation between $\sigma$ and~$k$ we can then find
$\Bsat$ as a function of~$k$ (figure~\ref{fig:bsat}, right).

A straightforward calculation shows that for small~$\vzn$ the peak
of~$\Bsat$ lies at wave number $\kpeak$ given by
\begin{equation}
    \kpeak^{2} c^{2} \simeq \ompe^2 + \ompp^2 / \gn^3 \; ,
    \label{eq:kpeak}
\end{equation}
with a corresponding growth rate
\begin{equation}
\label{grate}
	\sigma(\kpeak) \simeq \frac{\ompp}{\sqrt{2 \gn}} \:
	\left( \frac{\vxn}{c} \right) \; .
\end{equation}

The proton-driven Weibel instability occurs for a relatively small range in
wavelength compared to the electron-positron case (figure~\ref{fig:bsat}):
modes with wavelength longer than the electron skin depth ($k < c/\ompe$)
are inhibited by the response of the background electrons to the proton
perturbations.
The location where the growth rate levels off corresponds with the location
where the saturation magnetic field peaks ($k=\kpeak$).
If the background electrons had not been as responsive, this location might
have been at much lower wave number, and since the wave number is in the
denominator of expression~(\ref{eq:bsat}) for~$\Bsat$, this would have
resulted in a much higher magnetic field.
However, we find that the electrons set the location of the peak
($\kpeak\simeq\ompe/c$) and that the proton Weibel instability can only
produce slightly stronger magnetic fields than the electron instability
(figure~\ref{fig:bsat}, right), despite the larger kinetic energy of the
protons.

\section{Conclusion}

The Weibel instability as it was modeled here is not efficient at converting
the kinetic energy of the protons into magnetic fields.
However, numerical simulations of collisionless shocks in electron-proton
plasmas (\cite{fstm03}, \cite{fhhn03},~\cite{hasa03}) {\em do} show
efficient production of magnetic fields.
This can probably be attributed to merging of the electric currents
produced by the Weibel instability, {\em after\/} the instability that was
considered here has stopped.
To verify this, the current merging processes will need to be
investigated further.

The properties of the magnetic fields generated by these processes will be
important for determining the radiation that may be produced in these
shocks and that we see in the form of gamma-ray burst afterglows.

\bigskip\noindent{\bf Acknowledgment}: this research is supported by the
Netherlands Research School for Astronomy (NOVA).

\bibliographystyle{aipproc}
\bibliography{jw}

\begin{thebibliography}{9}
\expandafter\ifx\csname natexlab\endcsname\relax\def\natexlab#1{#1}\fi
\providecommand{\enquote}[1]{``#1''}
\expandafter\ifx\csname url\endcsname\relax
  \def\url#1{\texttt{#1}}\fi
\expandafter\ifx\csname urlprefix\endcsname\relax\def\urlprefix{URL }\fi

\bibitem[Rees and {M\'esz\'aros}(1992)]{reme92}
Rees, M.~J., and {M\'esz\'aros}, P., \emph{Mon. Not. Roy. Astron. Soc.},
  \textbf{258}, 41P (1992).

\bibitem[Gruzinov and Waxman(1999)]{grwa99}
Gruzinov, A., and Waxman, E., \emph{ApJ}, \textbf{511}, 852--861 (1999).

\bibitem[Kennel and Coroniti(1984)]{keco84}
Kennel, C.~F., and Coroniti, F.~V., \emph{ApJ}, \textbf{283}, 694--709 (1984).

\bibitem[Yang et~al.(1994)]{yal94}
Yang, T.-Y.~B., Arons, J., and Langdon, A.~B., \emph{Physics of Plasmas},
  \textbf{1}, 3059--3077 (1994).

\bibitem[Medvedev and Loeb(1999)]{melo99}
Medvedev, M.~V., and Loeb, A., \emph{ApJ}, \textbf{526}, 697--706 (1999).

\bibitem[Fonseca et~al.(2003)]{fstm03}
Fonseca, R.~A., Silva, L.~O., Tonge, J.~W., Mori, W.~B., and Dawson, J.~M.,
  \emph{Physics of Plasmas}, \textbf{10}, 1979--1984 (2003).

\bibitem[Frederiksen et~al.(2003)]{fhhn03}
Frederiksen, J.~T., Hededal, C.~B., Haugb{\o}lle, T., and Nordlund, {\AA}.,
  Magnetic field generation in collisionless shocks; pattern growth and
  transport, astro-ph/0308104 (2003).

\bibitem[Haruki and Sakai(2003)]{hasa03}
Haruki, T., and Sakai, J.-I., \emph{Physics of Plasmas}, \textbf{10}, 392--397
  (2003).

\bibitem[Silva et~al.(2002)]{sftm02}
Silva, L.~O., Fonseca, R.~A., Tonge, J.~W., Mori, W.~B., and Dawson, J.~M.,
  \emph{Physics of Plasmas}, \textbf{9}, 2458--2461 (2002).

\end{thebibliography}

\bigskip\noindent{\small
Copyright 2003 American Institute of Physics. This article may be
downloaded for personal~use only. Any other use requires prior permission
of the author and the American Institute of Physics.}

\end{document}